\documentclass[a4paper,conference]{IEEEtran}
\IEEEoverridecommandlockouts
\usepackage{amsmath,amssymb,amsfonts}
\usepackage{algorithmic}
\usepackage{graphicx}
\usepackage{textcomp}
\usepackage{xcolor}
\def\BibTeX{{\rm B\kern-.05em{\sc i\kern-.025em b}\kern-.08em
    T\kern-.1667em\lower.7ex\hbox{E}\kern-.125emX}}

\usepackage{multirow}
\usepackage{graphicx}
\usepackage{textcomp}
\usepackage{xcolor}
\usepackage{tabularx}
\usepackage{arydshln}
\usepackage{booktabs}
\usepackage{subcaption}
\usepackage{url}
\usepackage{hyperref}
\usepackage{mathtools}

\DeclarePairedDelimiter\floor{\lfloor}{\rfloor}

\begin{document}

\title{FIRST-SHOT ANOMALY SOUND DETECTION FOR MACHINE CONDITION MONITORING:\\ A DOMAIN GENERALIZATION BASELINE\\

}

\author{\IEEEauthorblockN{Noboru Harada, Daisuke Niizumi, Yasunori Ohishi, Daiki Takeuchi, and Masahiro Yasuda}
\textit{NTT Corporation}\\
Japan \\
E-mail: noboru.harada.pv@hco.ntt.co.jp}


\maketitle
\begin{abstract}
This paper provides a baseline system for First-shot-compliant unsupervised anomaly detection (ASD) for machine condition monitoring. First-shot ASD does not allow systems to do machine-type dependent hyperparameter tuning or tool ensembling based on the performance metric calculated with the grand truth. 
To show benchmark performance for First-shot ASD, this paper proposes an anomaly sound detection system that works on the domain generalization task in the Detection and Classification of Acoustic Scenes and Events (DCASE) 2022 Challenge Task 2: ``Unsupervised Anomalous Sound Detection for Machine Condition Monitoring Applying Domain Generalization Technique'' while complying with the First-shot requirements introduced in the DCASE 2023 Challenge Task 2 (DCASE2023T2).
A simple autoencoder based implementation combined with selective Mahalanobis metric is implemented as a baseline system. The performance evaluation is conducted to set the target benchmark for the forthcoming DCASE2023T2. Source code of the baseline system will be available on GitHub.
\end{abstract}


\section{Introduction}

This paper provides a baseline system for First-shot-compliant unsupervised anomaly sound detection (ASD) for machine condition monitoring.

Automatic machine condition monitoring with deep learning techniques for predictive maintenance is a crucial application in Industry 2.0/4.0.
In the Detection and Classification of Acoustic Scenes and Events (DCASE) Challenge, unsupervised anomalous sound detection tasks are held\cite{Koizumi_DCASE2020_01, Kawaguchi2021, Dohi2022-2}. However, most winning systems have utilized techniques specific to the challenge task setting.
In the task setting, many sound samples of similar but different machine instances are available as different sections of data for training. Some systems have used these different sections of data as pseudo anomaly data samples. 
Besides, most of the winning systems have relied on machine-type dependent hyperparameter tuning and tool ensembling based on the performance observation with the given anomaly samples for performance assessment. These solutions are not always applicable to the industry's realistic application scenarios. 

Since no restriction was specified in the task rule of the DCASE 2022 Challenge Task 2: Unsupervised Anomalous Sound Detection for Machine Condition Monitoring Applying Domain Generalization Techniques (DCASE2022T2), all of the top five winning systems \cite{LiuCQUPT2022, KuroyanagiNU-HDL2022, GuanHEU2022, DengTHU2022, VenkateshMERL2022} utilized techniques that made use of machine-type dependent hyperparameter tuning and tool ensembling for each machine type. Those systems were fine-tuned with oracle performance metrics, such as the AUC score calculated using the normal and anomaly samples provided for performance assessment. Therefore, the information from the anomaly samples had been leaked into the system.

Considering the above, the DCASE Challenge 2023 Task 2: First-shot Unsupervised Anomalous Sound Detection for Machine Condition Monitoring (DCASE2023T2) will introduce a First-shot requirement into the anomalous sound detection task. The First-shot ASD task does not allow systems to do any hyperparameter tuning referencing the performance metrics calculated with the grand truth, especially with anomaly samples provided for evaluating systems.

To show benchmark performance for the First-shot ASD, this paper provides an anomaly sound detection system that works on the domain generalization task in the DCASE2022T2\cite{Dohi2022-2} while complying with the First-shot requirements newly introduced in the DCASE2023T2. 

The DCASE2022T2 was for Domain Generalization. However, it can be interpreted as a highly data-unbalanced training task.
The DCASE2022T2 Domain Generalization task is a task that requires an efficient training scheme for unsupervised anomaly detection with highly unbalanced training data (990 normal samples from the source domain and only 10 samples from the target domain are given for training) and that no domain label be given for testing target samples. 

This paper proposes a simple implementation of Mahalanobis metric as the baseline system for the forthcomming DCASE2023T2 Challenge. This strategy is a tiny subset of what has been applied in Liu\_CQUPT\_task2\_4\cite{LiuCQUPT2022} which was the first-place winner of the DCASE2022T2.
The source code of the proposed baseline system will be made available on GitHub\cite{dcase2023t2ae}.

In complying with the First-shot requirement, the system is expected to perform similarly well on any real data to be seen in the industry.


\section{First-shot anomaly detection task}

\subsection{First-shot unsupervised ASD task}

Several efficient techniques were proposed in the previous DCASE ASD challenges for machine condition monitoring held in 2020, 2021, and 2022; some of the techniques made use of data from different section as anomaly samples. Besides, some of the techniques relied on machine-type dependent hyperparameter tuning and tool ensembling based on the performance observation with the given anomaly samples.

In general, it is hard to collect enough data sample variations from various machine instances and enough anomaly samples in real-world applications. Therefore, those solutions are unrealistic in such cases.

To encourage the challenge participants to provide truly useful technologies as real-world solutions, the First-shot requirement has been newly introduced for the forthcoming DCASE2023T2 \cite{dcase2023t2web}. 

The First-shot ASD is characterized as follows:
\begin{itemize}
\item No use of data from different sections (different machine instances) 
\item No hyperparameter tuning applied for dedicated machine type
\item No tool-ensemble applied for dedicated machine type
\end{itemize}

Instead of having statements describing the rule, the requirements are implicitly implemented in the task conditions so that compliant solutions will be naturally forced. For example, only a single section per each machine type is available; therefore, no classification technique is applicable. In addition, Machine types provided with the Evaluation datasets (to rank the systems) are completely different from the given machine types for the Development dataset; therefore, no hyperparameter tuning nor ensemble tool tuning by checking the grandtruth is possible.

\vspace{3mm}
\subsection{Domain generalization as a data unbalance task}
\label{domain_generalization}

The target task to solve in this paper is the DCASE2022T2: Domain Generalization task\cite{Dohi_arXiv2022_01, Harada_dcase2021, Dohi_arXiv2022_02} but with the task additionally restricted by the First-shot requirement.

\begin{figure}[tbp]
\vspace{5mm}
\centerline{\includegraphics[width=1.0\linewidth]{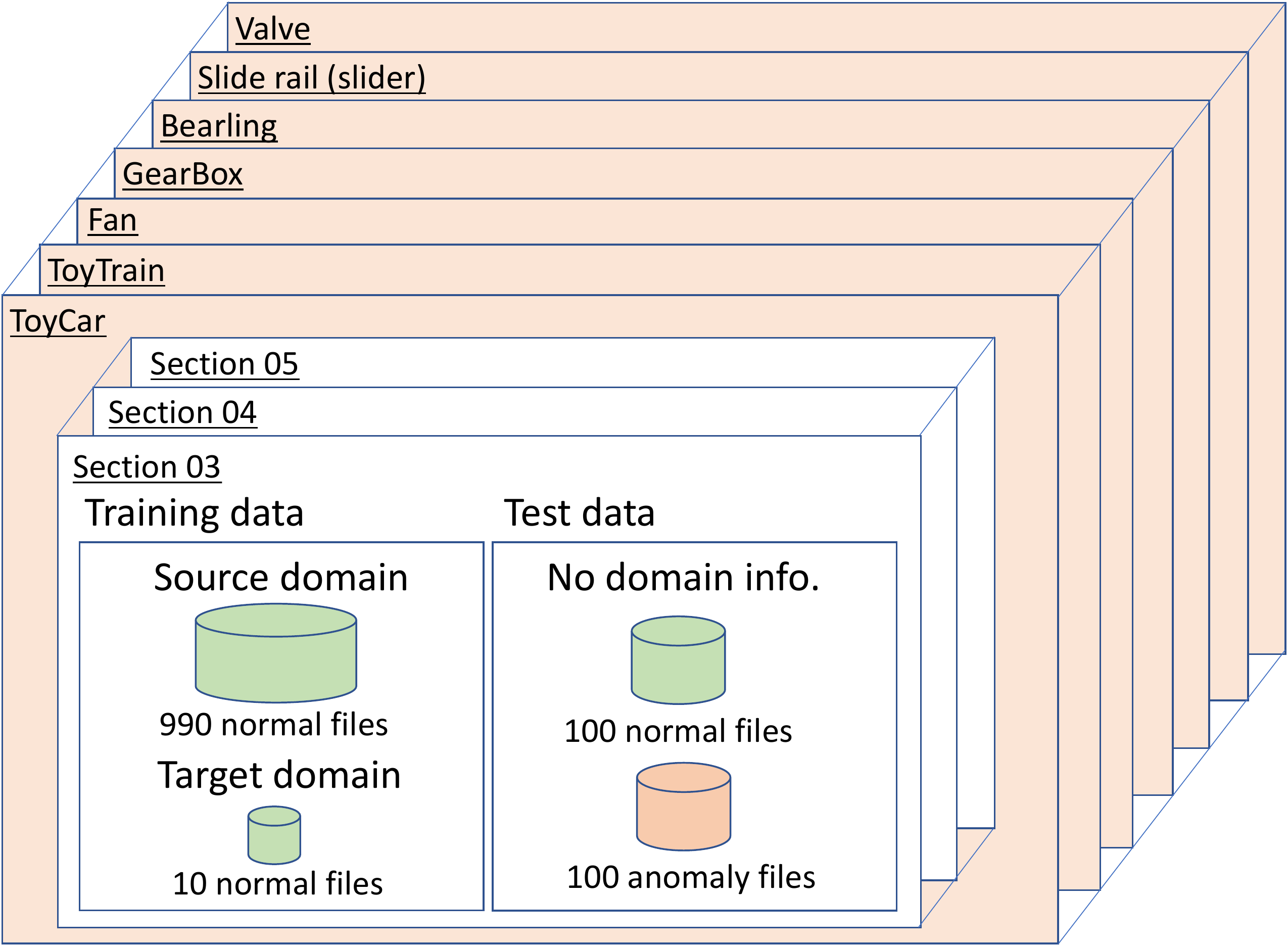}}
\vspace{3mm}
\caption{Additional training and Evaluation dataset of DCASE2022T2.}
\label{fig1}
\end{figure}

The Additional training and Evaluation datasets for the DCASE2022T2 evaluation were used. As shown in Fig.~\ref{fig1}, there are seven machine types; ToyCar, ToyTrain, Fan, Gearbox, Bearing, Slide rail (Slider), and Bearing. Each machine type has three sections reflecting the different operating conditions of the machines.
Each section has source and target domains representing domain shift conditions. Labels indicating source/target domains are given only for training. No domain label is given for testing.
There are 990 files given as training data for the source domain, and only 10 are given for training the target domain. All files contain 10 s of normal machine operating sound. No anomalous sound sample is given.
This task condition reflects an application scenario in which there are several operating modes but a limited number of recordings is available for training.
For testing systems, the Evaluation dataset of the DCASE2022T2 was used:
50 normal and 50 anomaly files for source and target domains; 200 unlabeled files per section in each machine type. 

\subsection{Performance evaluation metric}
\label{metrics}

Following the ranking rule of the DCASE2022T2, the total score $\Omega$ for evaluating the systems is calculated based on Area Under the Receiver Operating Characteristic (ROC) curve (AUC) and partial AUC (pAUC) with a harmonic mean ($hmean$) of AUC and pAUC using the following formula:

\begin{eqnarray}
\Omega &= hmean\{AUC_{m,n,d}, pAUC_{m,n}\\ \notag
  &|~~m \in \mathcal{M}, n \in \mathcal{S}(m), d \in \{source, target\} \},\label{eq1}
\end{eqnarray}
\begin{eqnarray}
AUC_{m,n,d}=\frac{1}{N_d^- N_n^+}\sum_{i=1}^{N_d^-}\sum_{j=1}^{N_n^+}\mathcal{H}(A_{\theta}(x_j^+)-A_{\theta}(x_i^-)),\label{eq2}\\
pAUC_{m,n}=\frac{1}{\floor*{pN_n^-}N_n^+}\sum_{i=1}^{\floor*{pN_n^-}}\sum_{j=1}^{N_n^+}\mathcal{H}(A_{\theta}(x_j^+)-A_{\theta}(x_i^-)).\label{eq3}
\end{eqnarray}
\vspace{5mm}

The AUC is calculated based on the anomaly scores $A_{\theta}(\cdot)$ of normal samples $x_i^-$ and anomaly samples $x_j^+$on each domain $d \in \{source, target\}$ of machine type $m \in \mathcal{M}$ and averaged over sections $n \in \mathcal{S}(m)$, where $m$ is the machine type indicator and $n$ is the section indicator in the machine type. 

The pAUC is calculated on each machine type $m \in \mathcal{M}$ and averaged over sections across domains. $N_d^-$ is the number of normal samples in each domain. $N_n^-$ and $N_n^+$ are the numbers of normal and anomaly samples in the section $n$ of the machine type $m$ respectively. $\mathcal{H}(\cdot)$ is a function that returns 1 when $x > 0$ and 0 otherwise. For the pAUC calculation, the false-positive rate (FPP) $p$ is set to 0.1. $\floor*{\cdot}$ is a floor function.


\begin{figure}[tbp]
\centerline{\includegraphics[width=1.00\linewidth]{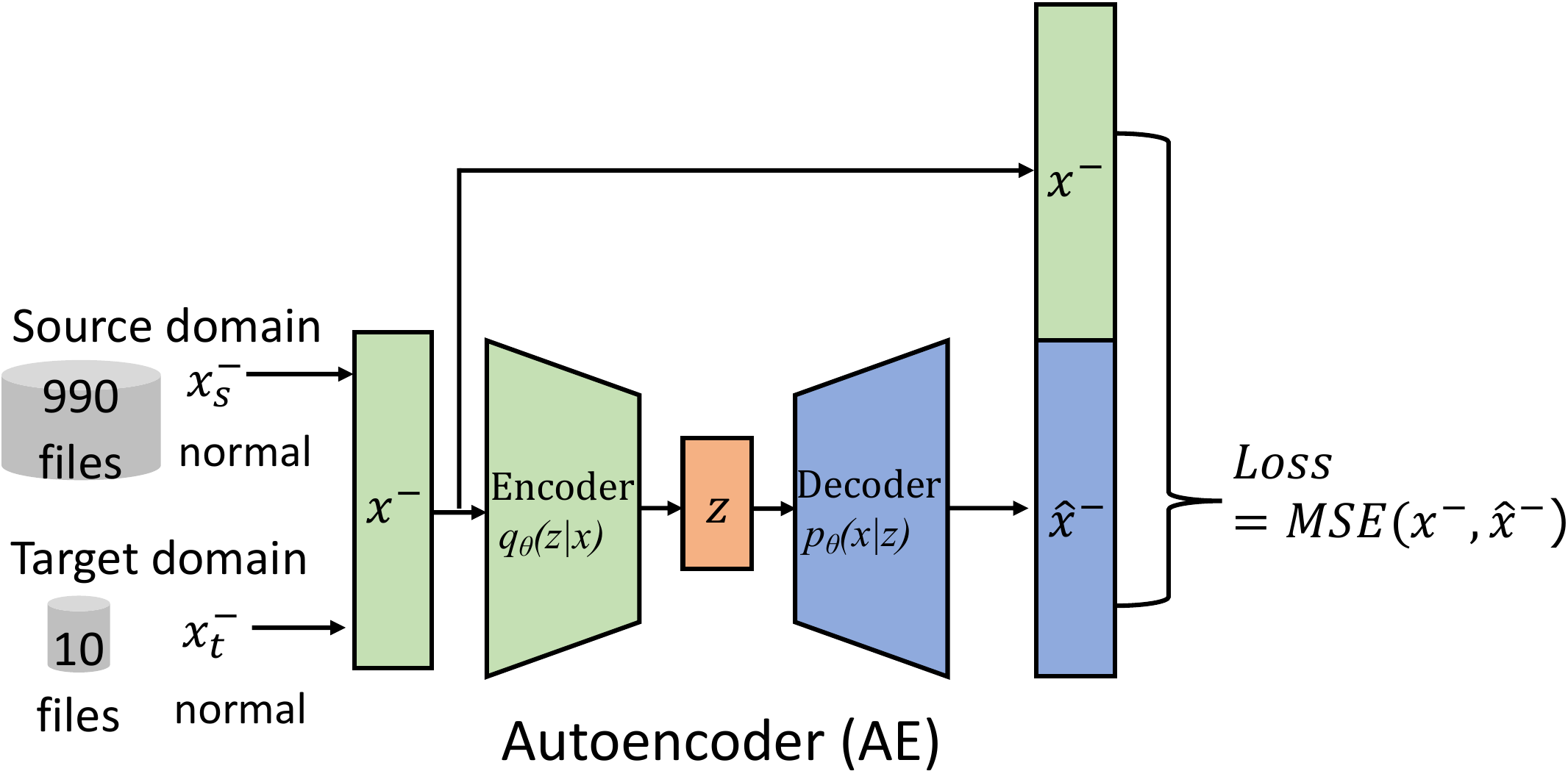}}
\caption{Overview of modified baseline Autoencoder.}
\label{fig2}
\vspace{-2mm}
\end{figure}

\newpage

\section{Proposed baseline systems for First-shot Domain Generalization}

Several systems were proposed in the DCASE2022T2.
Some of the winning systems adopted classification-based detectors.
Classification was performed among the samples from different sections given.
For example, there were three sections in the machine type ToyCar.
Note that there were two baseline systems in the DCASE2022T2; the DCASE 2022 task 2 Autoencoder (AE) and MovileNetV2 \cite{dcase2022t2ae, Sandler2018MobileNetV2IR, dcase2022t2MobileNetV2}. However, those two baseline systems are not compliant with the First-shot requirements.

This paper proposes a First-shot compliant baseline system that provide the target performance benchmark for the forthcoming DCASE2023T2.

\subsection{First-shot compliant baseline AE}
\label{baselineAE}

The First-shot compliant baseline AE is implemented based on the DCASE2022T2 baseline AE\cite{dcase2022t2ae} with slight modification. Its structure is shown in Fig.~\ref{fig2}.

While the DCASE2022T2 baseline AE is trained with all the section data in the target machine type, the First-shot ready version takes only the target normal samples from the dedicated section in the machine type. Therefore, three models dedicated to each section should be trained when there are three sections in the dataset.

Input audio samples are analyzed using STFT with 64-ms frames with 50\% hop and then converted into 128 bands of Log-mel energies. Five consecutive frames are concatenated to formulate input vectors to the autoencoder, which has 640 dimensions.
 
For training, the model parameter $\theta$ of the AE is trained to minimize the mean square error (MSE) between an input normal sample $x^{-}$ and its reconstruction $\hat{x}^{-}$ using
\vspace{-3pt}
\begin{eqnarray}
Loss & = MSE(x^-, \hat{x}^-),\label{eq4}\\
where &~~\hat{x}^- = Dec_{\theta}(Enc_{\theta}(x^-)).
\end{eqnarray}
For the testing phase, the anomaly score is calculated with the reconstruction error of the given query sample $x$ using
\begin{eqnarray}
Anomaly~Score~~A_{\theta} & = MSE(x, \hat{x}),\label{eq6}\\
where~~\hat{x} & = Dec_{\theta}(Enc_{\theta}(x)).
\end{eqnarray}
When the anomaly score exceeds the pre-set threshold, the sample is detected as an anomaly sample.

\subsection{Selective Mahalanobis AE}
\label{MahalanobisAE}
On top of the First-shot compliant baseline AE in \ref{baselineAE}, the Mahalanobis distance \cite{Mahalanobis1936} can be introduced to calculate the anomaly scores. This strategy is a tiny subset of what has been applied in Liu\_CQUPT\_task2\_4\cite{LiuCQUPT2022} which was the first-place winner of the DCASE2022T2.

The basic model structure and the training scheme are equivalent to the AE shown in \ref{baselineAE} except the covariance matrixes $\Sigma_{s}^{-1}$ and $\Sigma_{t}^{-1}$ of distance between normal samples $x^-$ and its reconstruction $\hat{x}^-$ for the source and target domains are calculated after the last epoch update of training.

The anomaly score $A_{\theta}$ is calculated using those covariance matrixes with:
\vspace{-2mm}
\begin{eqnarray}
Anomaly~Score~~A_{\theta} & = min\{D_s(x,\hat{x}), D_t(x,\hat{x})\},\label{eq8}\\
where~~D_s(\cdot) & = Mahalanobis(x,\hat{x},\Sigma_{s}^{-1}),\\
D_t (\cdot) & = Mahalanobis(x,\hat{x},\Sigma_{t}^{-1}).
\end{eqnarray}
We call this the Selective Mahalanobis AE in this paper.





\subsection{Design principle of the proposed baseline system}

Baseline models described in Secs. \ref{baselineAE} and \ref{MahalanobisAE} are implemented as one baseline source code having two operating modes. The new baseline is implemented with PyTorch while the previous DCASE2022T2 baseline autoencoder\cite{dcase2022t2ae} is implemented using Keras. The network architecture of those is equivalent.

There were several potential tools tested to be included. For example, tools introduced in \cite{Suefusa_ICASSP2020, Koizumi_WASPAA2019_02, Koizumi_IEEE2019} have been examined. However, the performance improvement of those tools was not significant enough in terms of the total score compared to the added complexity (i.e., the source code level complexity and the run-time computational complexity). 

Serving as a baseline implementation, simplicity is one of the most important aspects.
Therefore, we decided not to include them and rather keep it simple.

\begin{table*}[t!]
\caption{AUC results on the dataset from the DCASE 2022 Challenge Task 2 under domain-generalization conditions.}
  \label{tab:tab1}
  \centering
  \begin{tabular}{ l | c | c  c | c  c  c  c  c  c  c  } 
   \toprule
    \multicolumn{1}{c|}{System} & metric & hmean$^{*1}$ & amean$^{*1}$ & \textbf{ToyCar} & \textbf{ToyTrain} & \textbf{fan} & \textbf{gearbox} & \textbf{bearing} & \textbf{slider} & \textbf{valve} \\
    \midrule
    \textbf{First-shot compliant} & AUC (source) & 0.6484 & 0.6779 & 0.8229 & 0.4943 & 0.6609 & 0.7137 & 0.7475 & 0.7563 & 0.5493\\
    \textbf{baseline Autoencoder}  & AUC (target) & 0.5143 & 0.5451 & 0.6571 & 0.5195 & 0.3679 & 0.6116 & 0.5639 & 0.5275 & 0.5681\\
    \textbf{(FS-AE)} & pAUC (src \& tgt) & 0.5335 & 0.5407 & 0.6582 & 0.4967 & 0.5114 & 0.5200 & 0.5426 & 0.5361 & 0.5195\\
    random seed 13711 & \textbf{TOTAL score} & 0.5596 & 0.5879 & & & & & & & \\\hline					
    & AUC (source) & 0.6520 & 0.6822 & 0.8509 & 0.4986 & 0.6524 & 0.7087 & 0.7493 & 0.7624 & 0.5529\\
    & AUC (target) & 0.5143 & 0.5449 & 0.6778 & 0.5251 & 0.3575 & 0.5969 & 0.5627 & 0.5293 & 0.5653\\
    & pAUC (src \& tgt) & 0.5367 & 0.5452 & 0.6786 & 0.4988 & 0.5175 & 0.5128 & 0.5528 & 0.5347 & 0.5211\\
    random seed 13591 & \textbf{TOTAL score} & 0.5616 & 0.5908 & & & & & & & \\\hline				
    & AUC (source) & 0.6542 & 0.6808 & 0.8253 & 0.5063 & 0.6554 & 0.7125 & 0.7468 & 0.7595 & 0.5596\\
    & AUC (target) & 0.5084 & 0.5355 & 0.6263 & 0.5283 & 0.3487 & 0.5993 & 0.5597 & 0.5222 & 0.5639\\
    & pAUC (src \& tgt) & 0.5329 & 0.5402 & 0.6607 & 0.4965 & 0.5118 & 0.5128 & 0.5468 & 0.5356 & 0.5170\\
    brandom seed 13267 & \textbf{TOTAL score} &	0.5584 & 0.5855 & & & & & & & \\\hline					
    & AUC (source) & 0.6515 & 0.6803 & 0.8331 & 0.4998 & 0.6562 & 0.7116 & 0.7479 & 0.7594 & 0.5539\\ 
    & AUC (target) & 0.5123 & 0.5418 & 0.6537 & 0.5243 & 0.3580 & 0.6026 & 0.5621 & 0.5264 & 0.5657\\
    & pAUC (src \& tgt) & 0.5344 & 0.5420 & 0.6658 & 0.4973 & 0.5136 & 0.5152 & 0.5474 & 0.5355 & 0.5192\\
    \textbf{Average} & \textbf{TOTAL score} & \textbf{0.5599} & \textbf{0.5880} & & & & & & & \\\hline\hline
    \textbf{Selective Mahalanobis} & AUC (source) & 0.6597 & 0.7118 & 0.9528 & 0.4862 & 0.6789 & 0.8533 & 0.7216 & 0.7572 & 0.5329\\
    \textbf{AE} & AUC (target) & 0.5610 & 0.6082 & 0.8371 & 0.5043 & 0.4211 & 0.7801 & 0.6286 & 0.5568 & 0.5289\\
    (Tiny subset of Liu \cite{LiuCQUPT2022}) & pAUC (src \& tgt) & 0.5660 & 0.5805 & 0.7879 & 0.5118 & 0.5221 & 0.6104 & 0.5753 & 0.5389 & 0.5170\\
    random seed 13711 & \textbf{TOTAL score} & 0.5923 & 0.6335 & & & & & & & \\\hline					
    & AUC (source) & 0.6654 & 0.7140 & 0.9378 & 0.5048 & 0.6809 & 0.8493 & 0.7215 & 0.7660 & 0.5375\\
    & AUC (target) & 0.5533 & 0.5990 & 0.7977 & 0.5067 & 0.3969 & 0.7689 & 0.6300 & 0.5625 & 0.5301\\
    & pAUC (src \& tgt) & 0.5628 & 0.5779 & 0.7826 & 0.5035 & 0.5109 & 0.6095 & 0.5786 & 0.5440 & 0.5161\\
    random seed 13591 & \textbf{TOTAL score} & 0.5897 & 0.6303 & & & & & & & \\\hline
    & AUC (source) & 0.6700 & 0.7155 & 0.9297 & 0.5183 & 0.6862 & 0.8504 & 0.7153 & 0.7678 & 0.5409\\
    & AUC (target) & 0.5528 & 0.5924 & 0.7547 & 0.5116 & 0.3984 & 0.7681 & 0.6254 & 0.5593 & 0.5296\\
    & pAUC (src \& tgt) & 0.5580 & 0.5707 & 0.7509 & 0.5047 & 0.5068 & 0.6014 & 0.5751 & 0.5395 & 0.5161\\
    random seed 13267 & \textbf{TOTAL score} & 0.5890 & 0.6262 & & & & & & & \\\hline					
    & AUC (source) & 0.6650 & 0.7138 & 0.9401 & 0.5031 & 0.6820 & 0.8510 & 0.7195 & 0.7637 & 0.5371\\
    & AUC (target) & 0.5557 & 0.5999 & 0.7965 & 0.5075 & 0.4055 & 0.7724 & 0.6280 & 0.5595 & 0.5296\\
    & pAUC (src \& tgt) & 0.5623 & 0.5763 & 0.7738 & 0.5067 & 0.5133 & 0.6071 & 0.5763 & 0.5408 & 0.5164\\
    \textbf{Average} & \textbf{TOTAL score} & \textbf{0.5903} & \textbf{0.6300} & & & & & & & \\\hline\hline					
    
    \textbf{cf. DCASE2022T2} & AUC (source) & 0.6450 & 0.6755 & 0.7753 & 0.5981 & 0.6383 & 0.6975 & 0.7271 & 0.7605 & 0.5315\\
    \textbf{baseline AE$^{*2}$} & AUC (target) & 0.4515 & 0.4833 & 0.5100 & 0.3919 & 0.3267 & 0.5702 & 0.5384 & 0.4959 & 0.5501\\
    (not First-shot compliant) & pAUC (src \& tgt) & 0.5292 & 0.5356 & 0.6249 & 0.4944 & 0.4995 & 0.5209 & 0.5563 & 0.5391 & 0.5144\\
    & \textbf{TOTAL score} & 0.5305 & 0.5648 & & & & & & & \\\hline
    \textbf{cf. DCASE2022T2} & AUC (source) & 0.5907 & 0.6436 & 0.6347 & 0.4692 & 0.7088 & 0.6283 & 0.6067 & 0.7470 & 0.7103\\
    \textbf{baseline MobileNetV2}$^{*2}$ & AUC (target) & 0.4748 & 0.5230 & 0.4037 & 0.6302 & 0.3506 & 0.4619 & 0.5460 & 0.4971 & 0.7715\\
     (not First-shot compliant) & pAUC (src \& tgt) & 0.5358 & 0.5420 & 0.5582 & 0.5082 & 0.5191 & 0.4972 & 0.5091 & 0.5291 & 0.6726\\
     & \textbf{TOTAL score} & 0.5295 & 0.5695 & & & & & & & \\\hline
     \textbf{cf. DCASE2022T2 Top 1} & AUC (source) & 0.7734 & 0.8312 & 0.8818 & 0.6176 & 0.7435 & 0.9668 & 0.7190 & 0.9516 & 0.9380\\
     \textbf{Liu\_CQUPT\_task2\_4}$^{*2}$ & AUC (target) & 0.7212 & 0.7514 & 0.8851 & 0.7250 & 0.5484 & 0.8419 &	0.6827 & 0.7557 & 0.8213\\
     \cite{LiuCQUPT2022} (massive ensemble,& pAUC (src \& tgt) & 0.7464 & 0.7618 & 0.8845 & 0.7046 & 0.5734 & 0.8604 & 0.6885 & 0.7286 & 0.8387\\
     not First-shot compliant) & \textbf{TOTAL score} & \textbf{0.7097} & \textbf{0.7815} & & & & & & & \\
    \bottomrule
\multicolumn{10}{l}{$^{*1}$\textit{hmean} denotes harmonic mean, and \textit{amean} denotes arithmetic mean.  $^{*2}$The last three systems are not compliant with the First-shot requirements.}
  \end{tabular}
\end{table*}

\section{Experiments}

The DCASE2022T2 Additional training and Evaluation datasets described in Sec.~\ref{domain_generalization} were used to evaluate the performance of the proposed baseline systems.
The performance scores of the First-shot compliant baseline AE (FS-AE) and the Selective Mahalanobis AE were assessed with the performance evaluation metric described in Sec.~\ref{metrics}.

The frame size for STFT was 64 ms with 50\% hop size translated into 128 frequency bands Log-mel energies. Five consecutive frames were concatenated to formulate 640 dimensions (128 x 5) as input to the system. There were three layers of 128 dimensions linear, Batch normalization, and Activation with ReLU each. The bottleneck layer had eight dimensions. The number of epochs for training was 100.
The batch size was 256, and the Adam optimizer used a 0.001 learning ratio.



Test results with three different random seeds and the averaged scores are compared with scores of the DCASE2022T2 systems as shown in Table \ref{tab:tab1}.
The total score of the First-shot compliant baseline Autoencoder and the Selective Mahalanobis AE should be the target benchmarks for the DCASE2023T2 Challenge.


\section{Conclusion}

This paper proposed a baseline system for First-shot-compliant unsupervised anomaly detection scheme for machine condition monitoring.



The proposed system complies with the First-shot requirements; namely, it does not use any other data samples from different machine IDs, and no hyperparameter nor ensemble based on the performance observation of the grand truth of test data is used. It is trained with given normal training data only.

Experimental results showed that the performances of the proposed First-shot compliant baseline AE and the Selective Mahalanobis AE are better than the baselines of the DCASE2022T2 baseline AE and MobileNetV2.
It is also shown that the performances of those baseline AEs are not comparable to the first-place winner of the DCASE2022T2 that utilized the machine-type dependent hyperparameter tuning and massive tool ensemble.

Since the proposed system complies with the First-shot requirement, it is expected to perform similarly well on any real data in the industry.

The proposed system will be the baseline showing performance benchmark for the forthcoming DCASE 2023 Challenge Task 2: First-shot Unsupervised Anomalous Sound Detection for Machine Condition Monitoring. 
The source code of the proposed system is available on GitHub\cite{dcase2023t2ae}.

Further improvement should be expected throughout the Challenge.

\vspace{-2mm}
\section*{Acknowledgment}

The authors thank Mr. Yuri Musashijima for his hard work in implementing the systems and conducting experiments.

\newpage

\bibliographystyle{IEEEtran}
\bibliography{mybib}

\end{document}